\documentclass[aps,prx,twocolumn,superscriptaddress]{revtex4-2}

\usepackage{graphicx}
\usepackage{caption}
\usepackage{comment}
\usepackage{amsmath}
\usepackage{amssymb}
\usepackage{tabularx}
\usepackage{placeins}
\usepackage{units}
\usepackage{multirow}
\usepackage{braket}
\usepackage{color}
\usepackage{hyperref}
\usepackage[normalem]{ulem}
\usepackage{breakcites}
\usepackage{adjustbox}
\usepackage{mathtools}
\usepackage{float}
\usepackage{adjustbox}
\usepackage{dcolumn}
\usepackage{bm}

\usepackage{url}

\begin{document}

\title{ Quantum Monte Carlo
pair orbital wave functions for periodic systems towards the thermodynamical limit: ground states, excitations and spinors
}

\author{Lubos Mitas}

\affiliation{
 Department of Physics, North Carolina State University, Raleigh, North Carolina 27695-8202, USA\\
}

\date{\today}

\begin{abstract}
We derive 
 many-body single- and multi-reference wave functions for quantum Monte Carlo of periodic systems with an anti-symmetric portion that explicitly integrates over the Brillouin 
 zone of one-particle Bloch states. The wave functions are BCS-like determinants for singlets and pfaffians for polarized states built with appropriate pair orbitals. 
 This ab initio formalism is broadly applicable, eg, to description of quasi-particle band gaps, optical excitations and  to systems with complicated Fermi surfaces.  It generalizes to spin-dependent interactions using two-component spinor pairs.

\end{abstract}

\maketitle

Electronic structure calculations by many-body wave function methods have opened new opportunities for studies of advanced and complex materials. 
In particular, quantum Monte Carlo (QMC) methods have pushed
the boundaries of what is currently feasible in terms of system size, accuracy and variety of materials that were previously inaccessible computationally 
\cite{qmcrev,needsrev,lucasrev,KolorencMitasReview,sandrobook,can}. In real space particle configurations QMC methods are represented by the variational and diffusion Monte Carlo (VMC and DMC) approaches which have gained wider use because of growing availability of efficient computational tools. 
Despite many successes, further progress in calculations of more subtle quantum phenomena (magnetic effects, collective states, topological orders) is being hampered by two main sources of bias. The first one comes from the fixed-node approximation 
\cite{anderson76,reynolds82} used to circumvent fundamental difficulties that originate in varying signs 
of quantum amplitudes. In addition to this challenge, there is another source of errors which is encountered in calculations of condensed systems represented by supercells with periodicity. 
Extrapolations of expectation values to the thermodynamic limit (TDL) very often involve complications such as slow convergence, uneven behavior and sometimes even hard to resolve ambiguities. 
This is not unique to QMC since similar complications appear also in other correlated wave function methods\cite{andreas2016,garnet2017}. 

A number of techniques and corrections have been employed to overcome such finite size biases  for kinetic and potential energy components as well as for other  expectations \cite{fraser,andy99,paul99,markus2016,kwee,neil2020,neil2023}. For example, sampling of the Brillouin zone can be improved by choosing the most representative ${\bf k-}$point(s) \cite{sandro} or by repeated calculations with ${\bf k-}$point offset sampling and corresponding twisted boundary conditions \cite{twist}. 
In metals one encounters
  varying occupations for different sizes/shapes of supercells such that the grand canonical ensemble formulation has been proposed generalizing the ordinary canonical formalism
\cite{azadi}.
 Both potential and kinetic energy convergence can be improved by an ansatz for the static structure factor\cite{chiesa} or by employing corrections using Density Functional Theory (DFT) approaches\cite{kwee}. 
Other approaches include enforcing thermodynamic consistency in calculations of excited and ionized states with subsequently derived band gaps  \cite{cody,gani}.
In charged systems additional corrections based on dielectric properties have been applied, beyond just the simplest,
 charge compensation models \cite{yubo}. 
Despite a number of ideas and developed  methods, the elimination of finite size effects has often proved to be very costly and laborious with substantial effort still being invested into finding a more straightforward and robust solution. 



Here we outline theory and expressions for 
 construction of many-body wave functions for periodic systems with explicit integration of one-particle states over the Brillouin zone. This involves forming pair orbitals that capture the dominant one-particle contributions
 to the kinetic energy and other expectations, ie,
  the inhomogeneous factors of the Bloch orbitals. It leads to wave functions that are either BCS-like determinants (singlets) or pfaffians (all other cases). Straightforward generalizations include wave function expressions for calculations of quasi-particle band gaps from $(N-1)$-, $(N+1)$-electron systems, as well as for cases of optical excitations. The constructions offer opportunities to treat complicated Fermi surfaces as well as low-dimensional features on Fermi surfaces. It is also generalized to two-component spinor pair formulation for Hamiltonians with spin-dependent terms.

It appears that the integration of the Brillouin zone 
should be straightforward since it is routine in mainstream approaches such as DFT. Note that these methods typically work with reduced objects such as one-particle density that, due to the reduction, in many aspects already correspond to the TDL.
However, many-body methods based on a finite number of particles run into difficulties since  they rely on supercells and necessary extrapolations to reach the TDL. The first issue comes from the geometry of lattices and associated shell occupation effects.  Consider a non-interacting jellium in 3D (homogeneous electron gas) with a spherical Fermi surface of radius $k_F$ and a cubic grid of ${\bf k}$-points. The number of points inscribed inside the sphere with increasing grid resolution is the well-known Gauss circle/sphere problem. In the limit of the grid step $\Delta {\bf k}$ $\to 0$, the actual number of such points shows fluctuations that {\em grow} faster than $M^{1/3}$, where $M$ is the asymptotic value given by {\bf k}-point density $\times$ $4\pi k_F^3/3$ 
\cite{huxley,chamizo,heath}. 
Although for intensive quantities (eg, cohesion per particle) this effect can be eventually suppressed by sufficiently large supercells  \cite{gani,cody}, the differences of total energies or other extensive quantities can be affected very substantially. Note that for growing simulation sizes this intermingles with an increase in statistical fluctuations characterized by the variance that is proportional to the number of treated particles. 
The shell effect impacts many properties, for example, sampling of complicated Fermi surfaces (FS) since one can have multiple sheets, pockets or spectral peaks which make convergence to TDL challenging. Another step in difficulties comes from lower-dimensional features on FS, eg arcs or Dirac/Weyl points.
 Although the twist averaging or grand canonical ensemble techniques should alleviate this, biases and slow convergence are difficult to dismiss. It is therefore highly desirable to construct wave functions that would integrate over the Brillouin zone states from the outset. 


%
%
%
%


{\em Singlets. } Let us consider a singlet state with $N$ electrons and the commonly used Slater-Jastrow trial function given by an antisymmetric part and the Jastrow correlation factor
\begin{equation}
\Psi_T=\Psi_Ae^{J({\{i\},\{j\}})} 
={\rm det}[\phi_k^{\uparrow}(i)]
{\rm det}[\phi_k^{\downarrow}(j)]e^{J({\{i\},\{j\}})},
\end{equation}
where Slater matrices have the dimensions
 $(N/2)\times (N/2)$,
 $k$ labels one-particle states and $i,j$ denote
electron positions ${\bf r}_i,{\bf r}_j$. (Throughout the narrative we focus only on the antisymmetric part and we drop any normalizations that are not essential.)  
The antisymmetric component can be rewritten as
\begin{equation}
\Psi_A=\det[\Phi(i,j)], \;\,
{\rm where}\;\;  \Phi(i,j)=\sum_{k=1}^{N/2}\phi^{\uparrow}_k(i)\phi^{\downarrow}_k(j)
\label{eqn:bcs}
\end{equation}
is  a
 singlet pair orbital. In order to capture the thermodynamic limit for periodic systems we expand the sum to the full integral over the Brillouin zone (BZ). A straightforward illustration is jellium with $0\leq |{\bf k}|\leq k_F$ and Hartree-Fock (HF) eigenstates $e^{i{\bf k}\cdot{\bf r}}$, $e^{-i({\bf k}\cdot{\bf r})}$. 
We obtain
\begin{equation}
\label{eqn:hfcos}
\Phi(i,j)=\int[e^{i{\bf k}\cdot{\bf r}_{ij}}+ e^{-i{\bf k}\cdot{\bf r}_{ij}}]d{\bf k}=2\int 
\cos({\bf k}\cdot {\bf r}_{ij}) d{\bf k}  
\end{equation}
where ${\bf r}_{ij}={\bf r}_i-{\bf r}_j$. Therefore $\Psi_A$ for jellium is given by 
\begin{equation}
\Psi_A^{jell}=\Psi_{ HF}^{jell} = {\rm det}[{j_1(k_Fr_{ij})/ (k_Fr_{ij})}]
\end{equation}
where $r_{ij}=|{\bf r}_{ij}|$,
while
$j_1(.)$ is the spherical Bessel function. 
 This remarkably compact expression reminds us that the pair orbital is related to the amplitude of 
 the HF single-particle density matrix component as it becomes clear from substitutions ${\bf r}_i\leftarrow {\bf r}, {\bf r}_j\leftarrow {\bf r}'$ \cite{ziesche}. 
  For inhomogeneous systems with ionic potentials and corresponding one-particle Bloch orbitals $e^{i{\bf k}\cdot {\bf r}} u_{n,{\bf k}}({\bf r})$, $e^{-i{\bf k}\cdot {\bf r}}u_{n,-{\bf k}}({\bf r})$ the pair orbital is given by
  \begin{equation}
  \label{eqn:singlet}
  \Phi(i,j)=\sum_n\int [e^{i{\bf k}\cdot{\bf r}_{ij}}u_{n,{\bf k}}({\bf r}_i)
  u_{n,-{\bf k}}({\bf r}_j)+ c.c.] d{\bf k}
  \end{equation}
   where $n$ is  the band label and $c.c.$ is the complex conjugated term.  So far both the summation and the integration involved only occupied states. We expand the variational freedom beyond HF by including both occupied and
     virtual states with corresponding 
 coefficients  
\begin{equation}
\Phi(i,j)=\sum_n\int c_n({\bf k})[...] d{\bf k} 
\end{equation}
where the square bracket repeats from Eq.(\ref{eqn:singlet}). 
 The determinant of this pair orbital (as in Eq.(\ref{eqn:bcs})) is the BCS-like wave function (ie, Bardeen-Cooper-Schrieffer state projected on the fixed number of particles).
 For simplicity here we assume diagonalized pair function expanded in natural orbitals with correpsonding coefficients. 
 Generalizations to broken symmetries with spin unrestricted orbitals and off-diagonal terms are elaborated in a broader context below.

{\em Finite supercell.} We need to consider that the presented expressions nominally correspond to an infinite system.
 However, in actual simulations our systems will be finite and we have to adapt the construction accordingly. The first obvious point is
that
the derived wave functions are not periodic for any finite supercell size.
In order to address this, it is useful to recall 
that periodicity limits genuinely independent information about inter-particle correlations up to the minimum image distance.
We also note that the non-periodicity concerns only the plane wave factors in Eqs.(\ref{eqn:hfcos}),(\ref{eqn:singlet}). Taking these two considerations into account, we enforce periodicity of the wave function by imposing supercell minimum image distances in the plane wave factors  
\begin{equation}
{\bf r}_{ij} \leftarrow \tilde{{\bf r}}_{ij} ; \qquad\qquad
|\tilde{{\bf r}}_{ij}|= {\rm min}_{\bf S} |{\bf r}_{ij}+{\bf S}|
\end{equation}
where ${\bf S}$ represents the supercell lattice vectors. Fortunately, the analytical structure of this restriction is very transparent such that one can correct (eg, smooth out) wave function derivative discontinuity from rewinding or even directly include long-range corrections through a variational ansatz. This is a minor overhead considering  
that we have integrated the orbital inhomogeneous factors that can be almost arbitrarily complicated. 
Note that a related restriction takes place in ordinary QMC calculations as well since particle positions are limited (or rewound) to the simulation supercell.
We recall that for the potential energy the periodicity of Coulomb interactions is treated by Ewald sums that result in bare Coulomb interactions within the minimum image distance compensated by size-dependent periodicity corrections. 

{\em Spin polarized states.} 
For fully spin polarized systems the pair orbital has to possess the triplet symmetry. In analogy with Eq.(\ref{eqn:singlet}) this is given
by
\begin{equation}
\label{eqn:triplet}
{ \xi}^{\uparrow\uparrow}(i,j)=\sum_{k=1,3, ...}^{N-1} \varphi_k(i)\varphi_{k+1}(j)
-
\varphi_k(j)\varphi_{k+1}(i)
\end{equation}
where we assume that $N$ is even (the case of $N$ odd is treated below). We can follow similar steps as in the case of a singlet pair construction 
and its generalization with the Bloch orbitals. Note that the integration domain involves only half of the Brillouin zone since 
the pair includes both 
 ${\bf k}$ and $-{\bf k}$ states and the integrated function is odd. 
Once this is constructed the wave function is formed by the 
antisymmetrization
of the product of $N/2$ triplet pairs. That results in a pfaffian
\begin{equation}
\label{eqn:triplet2}
\Psi_A={\cal A} \prod_{i=1,3 ...}^{N-1} { \xi}^{\uparrow\uparrow}(i,i+1)={\rm pf} [{ \xi}(i,j)]
\end{equation}
where the corresponding matrix of dimension $N$ is skew symmetric.

{\em Partially spin-polarized states and fundamental gap.} 
Systems with partial polarization are naturally described by the wave function that we introduced some time ago \cite{Bajdich2006,Bajdich2007}. The pfaffian of a matrix with singlet, triplet pairs and unpaired orbitals is given by 
\begin{align}
\Psi_{A}=
{\rm pf}\begin{bmatrix}
{\boldsymbol \xi}^{\uparrow\uparrow} &
{\boldsymbol \Phi}^{\uparrow\downarrow} &
{\boldsymbol\varphi}^{\uparrow} \\
-{\boldsymbol \Phi}^{\uparrow\downarrow T} &
{\boldsymbol \xi}^{\downarrow\downarrow} &
{\boldsymbol \varphi}^{\downarrow} \\
-{\boldsymbol\varphi}^{\uparrow T} &
-{\boldsymbol\varphi}^{\downarrow T} &
0 \;\; \\
\end{bmatrix},
\end{align}
where ${\boldsymbol \xi}^{\uparrow\uparrow}, 
{\boldsymbol \xi}^{\downarrow\downarrow},{\boldsymbol \Phi}^{\uparrow\downarrow}$ are block matrices and 
${\boldsymbol \varphi}^{\uparrow},{\boldsymbol \varphi}^{\downarrow}$ are vectors of unpaired orbitals. The unpaired channel is included for cases of  $N$ being odd 
 as otherwise the pfaffian vanishes. 
All the forms presented so far are obviously special cases of this general wave function.

Let us now consider 
the quasi-particle band gap calculated by the familiar expression as the difference between the neutral system and systems with an added or subtracted electron. 
As a toy illustration we assume
 two valence electrons in a singlet pair $\Phi(i,j)=\varphi_v(i)\varphi_v(j)$ plus a third ("$N+1$") electron occupying a conduction state $\varphi_c$.
  The simplest triplet pair function is then
\begin{equation}
\xi^{\uparrow\uparrow}(i,j)=\varphi_v(i)\varphi_c(j)-\varphi_v(j)\varphi_c(i),
\end{equation}
so that $\Psi_A$ above is given by 
\begin{equation}
\label{eqn:stu}
{\rm pf}\begin{bmatrix}
0 &
 \xi^{\uparrow\uparrow}(1,2)  &\Phi^{\uparrow\downarrow}(1,3)&
\varphi^{\uparrow}_{c}(1) 
\\
 \xi^{\uparrow\uparrow}(2,1) &
0&\Phi^{\uparrow\downarrow}(2,3) &
\varphi^{\uparrow}_{c}(2)
 \\
-\Phi^{\uparrow\downarrow}(3,1) &-\Phi^{\uparrow\downarrow}(3,2)
 & 0 &  \varphi_{v}^{\downarrow}(3) \\
-\varphi^{\uparrow}_{c}(1)
 &
 -\varphi^{\uparrow}_{c}(2)
 &
-\varphi_{v}^{\downarrow}(3) & 0\\
\end{bmatrix} \propto \Psi_{HF}.
\end{equation}
Since we simplified the pair functions to the minimal occupation,  we recover the Hartree-Fock result. It is perhaps less obvious that the same is also obtained for
cases where {\em either} the singlet {\em or} the triplet pair function is zeroed out. There is an
optional variational flexibility that can be explored by independent optimizations of singlet and triplet
channels \cite{Bajdich2006}. 
The same wave function also describes a similar toy illustration for a positively charged 
system. For this case we initially assume four electrons in a singlet state with two valence orbitals  $\varphi_{v}, \varphi_{v2}$ and $\Phi(i,j)=\varphi_{v}(i)\varphi_{v}(j)+\varphi_{v2}(i)\varphi_{v_2}(j)$. 
 The corresponding $(N-1)$-electron wave function is obtained  by 
 relabeling 
$\varphi_{c} \leftarrow \varphi_{v2}$ in Eq.(\ref {eqn:stu}).
The generalizations for arbitrary $N$, different spin states and the inclusion of virtual space are all straightforward, see also Refs.\cite{Bajdich2006, Bajdich2007,acta}. 

\medskip

{\em Optical/excitonic gap.} In calculations of optical excitations with fixed number of particles, QMC trial functions usually employ one-particle promotion from a valence to conduction band orbital,  $\varphi_v\to\varphi_c$. 
For insulators with large dielectric constants the corresponding excitonic effect is typically very small due to the delocalized nature of such Wannier excitons.
Obviously, then the fundamental and promotion gaps should be very close and indeed this has been true for a diverse set of materials  such as Si crystal\cite{gani}, 2D phosphorene\cite{tobi},  perovskite LaScO$_3$\cite{cody}, etc. 
The other limit of localized and strongly bonded Frenkel exciton typically entails significant restructuring of the wave function which almost inevitably results in a multi-reference character.  So far in QMC such cases have not been fully explored, mostly due to technical challenges such as the complicated character of the wave function construction and optimization \cite{fluorogra}. 

Fortunately, the pfaffian wave function offers possibilities of describing excitations beyond a single state promotion. In order to illustrate this we first remind
that once an integral over ${\bf k}$-space is involved, Eq.(\ref{eqn:singlet}),
promotion of a single orbital results in a zero measure effect. Indeed, we have to boost the excitation amplitude by a finite measure contribution, ie, by introducing a finite interval of states in the ${\bf k}$-space ("linewidth") to be included into the integral.   
For a singlet state electron-hole excitation the corresponding pair function is then modified as follows
\begin{equation}
\label{eqn:gvgc}
\Phi =\int g_v({\bf k},{\bf k}')[\varphi_{v{\bf k}}\varphi_{v{\bf k'}} +
\varphi_{v{\bf k}}\varphi_{v{\bf k'}}
] d{\bf k}d{\bf k}' +
\nonumber
\end{equation}
\begin{equation}
+\int g_{c}({\bf k},{\bf k}')[\varphi_{c{\bf k}}\varphi_{v{\bf k'}} +
\varphi_{c{\bf k}}\varphi_{v{\bf k'}}
] d{\bf k} d{\bf k'}
\end{equation}
where the amplitudes of valence and conduction states are represented by the functions $g_v$ and $g_{c}$. For completeness we included off-diagonal terms and we also split the coefficients into valence and conduction components. This split has its role since for the definition of an electron-hole pair there is a constraint for the conduction/valence ratio of norms to be $1/(N-1)$. 
Clearly, the functions $g_v$ and $g_{c}$ should be optimized to reflect both the exciton formation and relaxation of the valence band with a hole. 
With increasing supercell size these functions will evolve towards a more narrow linewidth and one expects monotonous trends towards the TDL. Note that the form enables restructuring of the wave function within the variational freedom provided by the single BCS-like determinant or pfaffian. However, this variational freedom might not be sufficient such that it is useful to consider  multi-reference expansions\cite{Bajdich2007}. 

{\em Multi-reference expansions.}
The single-pfaffian wave function is defined, in general, by the orbital set ${\boldsymbol \Omega}=$ 
$\{{\boldsymbol \xi}^{\uparrow\uparrow}, 
{\boldsymbol \xi}^{\downarrow\downarrow},{\boldsymbol \Phi}^{\uparrow\downarrow},
{\boldsymbol \varphi}^{\uparrow},{\boldsymbol \varphi}^{\downarrow}\}$, Eq. (\ref{eqn:stu}). However, the variational freedom can be further expanded by adding 
 another configuration(s). Although this seemingly appears similar to expansions in determinants, there are differences. For example, let us construct a two-configuration wave function with an excited pair orbital ${\Phi}^{ex}(i,j)$ that targets an improved description of unlike spin pair correlations. Such a two-configuration function is given by
\begin{equation}
\Psi_{A}={\rm pf}[{\boldsymbol\Omega}]+\sum_l {\rm pf}[{\boldsymbol\Omega};\Phi(l,.),\Phi(.,l)\leftarrow \Phi^{ex}(l,.), \Phi^{ex}(.,l)]
\end{equation}
where the pair orbital $\Phi$ in $l$-th row {\em and} column is replaced by $\Phi^{ex}$.  
 The additional reference 
is therefore composed from ${\cal O}(N)$ pfaffians/determinants in order to guarantee the antisymmetry.  ${\Phi}^{ex}(i,j)$ is defined by independent functions $g_v^{ex},g_{c}^{ex}$ (see Eq.(\ref{eqn:gvgc})) that target the sought after correlations. 

Generalization to $n$ independent pair function excitations
requires ${\cal O}(N^n)$
pfaffians/determinants, ie, for small $n$ such wave functions maintain polynomial scaling and therefore they are computationally feasible. There are many directions where such expansion can be effective: descriptions of exciton-exciton interactions, capturing excitations in quantum condensates as well as for description of lower-dimensional FS features such as arcs, Dirac points, etc. It is understood
that pair orbitals for correlating, say, low-D FS feature will require amplitudes with finite measure in ${\bf k}-$space in analogy to the case of excitons described above.

{\em Spinor pair orbitals.}
The presented formalism is straightforward to generalize to Hamiltonians with explicit spin dependencies. One-particle states are two-component spinors given by \cite{cody2016, cody2017}
\begin{equation}\chi(i)=
\chi({\bf r}_i,s_i)=
[a\chi^{\uparrow}(s_i)\varphi^{\uparrow}({\bf r}_i) +
b\chi^{\downarrow}(s_i)\varphi^{\downarrow}({\bf r}_i)].
\end{equation}
The antisymmetric spinor pair function is constructed from a Kramers' pair of spinors with labels $\alpha,-\alpha$ as follows
\begin{equation} 
X(i,j)=\sum_{n,\alpha}\int [\chi_{n,\bf k}^{\alpha}(i)\chi^{-\alpha}_{n,-\bf k}(j)-\chi_{n,\bf k}^{\alpha}(j)
\chi_{n,-\bf k}^{-\alpha}(i)] d{\bf k}.
\end{equation}
 Here the Bloch states are given by 
\begin{equation}
\chi_{n,\bf k}^{\alpha}(i)=e^{i{\bf k}\cdot{\bf r}_i}u^{\alpha}_{n,{\bf k}} ({\bf r}_i,s_i)   
\end{equation}
where $\alpha=\pm 1$ labels the two spinors $u^{\alpha}_{n,{\bf k}}(.)$ for a given $n,{\bf k}$. 
The trial function is simply written as 
\begin{equation}
\Psi_A
= {\rm pf} [X({\bf r}_i,s_i,{\bf r}_j,s_j)]
\end{equation}
where it is tacitly assumed that for odd $N$ the skew-symmetric matrix is boosted by an
additional unpaired spinor row and column.
The resulting pair orbital is an explicit mix of singlet and triplet pairs as expected for Hamiltonians with spin terms, eg, spin-orbit. For use of such wave functions in QMC, the fixed-node approximation is generalized into the fixed-phase method\cite{cody2017,ortiz}.

{\em Discussion and conclusions.} 
The QMC calculation's cost is determined by fluctuations of sampled quantities such as local energy and demands on evaluating the trial function. The fluctuations  approximately correspond to the average variance that is observed, for example, in twist sampling. For a given set of bands throughout the Brillouin zone this varies only marginally. 
Note that with increasing size this contribution to the variance diminishes and already for medium supercell sizes the overall accuracy of the correlation description (ie, quality of Jastrow factor, nodal errors, etc) becomes dominant.
Finally, we want to briefly mention the scaling with respect to the number of particles. The crucial consideration is the efficient evaluation of pair orbitals and the most obvious option is to use the diagonalized form that leads to the same scaling as 
 calculations with Slater determinants. (Another option is 6D interpolation, eg, by a product of lower-dimension splines, with the same type of scaling albeit with different prefactors.) The overall scaling is therefore the same as in ordinary variational and diffusion QMC calculations.

Let us  summarize the key points. 
By integrating over the Brillouin zone of one-particle 
Bloch orbitals we have constructed variational/trial wave functions that avoid ${\bf k}-$point sampling and associated uncertainties that complicate reaching the  thermodynamic limit. 
The resulting BCS-like determinants or pfaffians are applicable to periodic systems in general, including  metals with complicated Fermi surfaces. In addition, this allows for variational optimization of Fermi surfaces in the presence of Jastrow factors in an efficient and direct manner. For insulators we have outlined generalizations for evaluations of fundamental band gaps and optical excitations. 
The presented formalism can be adapted for systems with low-dimensional features on the Fermi surface such as Dirac
or Weyl points,
arcs, etc, therefore opening new paths for studies of systems with complex electronic phase diagrams. 
Both BCS-like determinants and pfaffians offer variational freedom beyond Slater determinants since pair orbitals couple electron up- and down-spin spaces with possibilities for gains in the fixed-node/phase quality and further systematic improvements using multi-reference expansions \cite{Bajdich2006}. The generalization to spin-dependent Hamiltonians is straightforward through construction of pfaffians of spinor pairs.
The formalism can describe collective states such as fermionic condensates in ultracold atomic gases  
\cite{xin,kevin}. These advances open new perspectives for finite size calculations of periodic systems by quantum Monte Carlo and other many-body wave function approaches. 
\bigskip
\begin{acknowledgements}

We thank Benjamin Kincaid and Paul R. C. Kent for reading the manuscript and for providing helpful suggestions.

This work has been supported by the U.S. Department of Energy, Office of Science, Basic Energy Sciences, Materials Sciences and Engineering Division, as part of the Computational Materials Sciences Program and Center for Predictive Simulation of Functional Materials.

\end{acknowledgements}

\bibliography{main.bib}

\end{document}